# Thermal Effects on Damping Determination of Perpendicular MRAM Devices by Spin-Torque Ferromagnetic Resonance

*H.J. Richter, G. Mihajlović, R.V. Chopdekar, W. Jung, J. Gibbons, N.D. Melendez, M.K. Grobis and T.S. Santos*

*Western Digital Research Center, Western Digital Corporation, San Jose, California 95119, USA*

**Abstract:** *We report device level damping measurements using spin-torque driven ferromagnetic resonance on perpendicular magnetic random-access memory (MRAM) cells. It is shown that thermal agitation enhances the apparent damping for cells smaller than about 55 nm. The effect is fundamental and does not reflect a true damping increase. In addition to the thermal effect, it is still found that device level damping is higher than film level damping and increases with decreasing cell size. This is attributed to edge damage caused by device patterning.*

## I. INTRODUCTION

Magnetic random-access memory (MRAM) is an attractive technology for high performance non-volatile memory. This technology makes use of the spin-transfer-torque (STT) effect such that the magnetic memory cells can be switched by an electrical current. To achieve high density, the memory elements should have their easy axes oriented perpendicularly. Perpendicular MRAM cells consist of a reference layer (typically the reference layer is a part of a synthetic antiferromagnetic structure), and a free layer separated by a tunneling barrier based on MgO, see e.g.,[1]. For switching, the Gilbert damping $\alpha$ of the free layer plays a key role. The spin-torque effect acts as negative damping and pumps energy into the magnetic system with each precession cycle[2]. The critical current for STT switching is given by[1,2,3]:

$$I_{c0} = \frac{\alpha}{\eta} \frac{2e_0}{\hbar} \mu_0 H_A' M_s V$$

(1)

Here, $\mu_0 = 4\pi \times 10^{-7}$ Vs/Am, $e_0$ is the elementary charge, $\eta$ the spin polarization efficiency $\eta = \frac{\sqrt{TMR(TMR+2)}}{2(TMR+1)}$ with TMR = tunneling magnetoresistive ratio, $\hbar$ the reduced Planck's constant, $M_s$ the saturation magnetization, $H_A'$ the total anisotropy field, and $V$ the volume of the free layer. Equation (1) holds for a macrospin, which means that the exchange forces are so strong that no inhomogeneous magnetization can occur. In this case, it is convenient to combine the perpendicular anisotropy field $H_A$ with that of the shape anisotropy $H_A' = H_A + \frac{M_s}{2}(N_\perp - N_\parallel)$ where $N_\perp, N_\parallel$ are the demagnetization factors perpendicular and parallel to the easy axis, respectively. It is understood that the concept of an anisotropy field applies to bulk materials whilst the origin of the perpendicular anisotropy in the free layer of the MRAM is a surface anisotropy originating from the interfaces between the free layer and the MgO[4]. Owing to the small thickness of MRAM cells, it is justified to use the concept of an anisotropy field.

Equation (1) shows that small damping is favorable to lower the switching current, and for the cells investigated here we measured the damping for the materials used in the free layers. These measurements are usually done by ferromagnetic resonance (FMR) on the film level, but not on the device level. Owing to the patterning process it may be expected that the device level damping is higher than the film level damping[5]. Here we investigate device level damping measured by spin-torque driven ferromagnetic resonance (STFMR). The paper is organized as follows: section II gives experimental details about the samples and the measurement procedure, section III outlines a basic theory for STFMR, and section IV reports on damping measurements on devices down to 20 nm in size. In section V, the effect of thermal agitation is shown to have a profound effect on the apparent damping for small MRAM cells. In section VI the interpretation of device level damping data is discussed.

## II EXPERIMENTAL

### II.1 Samples

The MRAM film stacks used in this study consist of a seed layer, a synthetic antiferromagnet, a reference layer, a MgO tunnel barrier and a CoFeB-based free layer. In addition, we employ a MgO cap layer to enhance anisotropy. Further details were discussed in references[6,7,8,9] and are not repeated here, but the relevant magnetic data are replicated in Table I.

| Free layer | $M_s$ (kA/m) | $H_A - M_s$ (kA/m) | $\delta$ (nm) | $A_{ex}$ (pJ/m) | $RA$ ($\Omega\mu m^2$) | $TMR$ (%) | $\alpha_0$ |
|---|---|---|---|---|---|---|---|
| A | 1276 | 232 | 2.05 | 11.3 | 6.4 | 87 | 0.0064 |
| B | 1034 | 166 | 2.17 | 8.92 | 11.1 | 109 | 0.011 |
| C | 1515 | 375 | 1.44 | 18.5 | 8.2 | 150 | 0.0038 |

**TABLE I**: *Various relevant properties of films of the three free layer materials investigated. Saturation magnetization $M_s$ [(i)], $H_A - M_s$ [(ii)], thickness $\delta$, resistance area product $RA$ [(iii)], $TMR$ [(iii)], and damping $\alpha_0$ [(ii)]. [(i)] measured by vibrating sample magnetometry, [(ii)] by ferromagnetic resonance of the un-patterned film, [(iii)] Inferred from device-level measurements as outlined in Refs. 7 and 8. $A_{ex}$ is the inferred exchange stiffness, see Ref .7.*

### II.2 STFMR Measurements

For the determination of device level damping, we have chosen STFMR because it is the only technique that can provide a high enough sensitivity to successfully characterize small MRAM devices [10,11,12,13]. In STFMR, the resonance is excited by an electrical current rather than a magnetic field as in conventional FMR. MRAM cells with perpendicular anisotropy are of greatest technological interest because they allow high density and lower switching current compared to cells with in-plane anisotropy. Although STFMR is well known for the characterization of the dynamic behavior of magnetic tunnel junction devices, only very few works on STFMR measurements on such cells have been reported [14,15,16,17]. In the perpendicular geometry, the easy axes of the free and the reference layers coincide with the $z$-axis. Hence, the resulting spin-transfer torque $\boldsymbol{m} \times \boldsymbol{m_p}$ is zero and no resonance can be excited ($\boldsymbol{m}$: magnetization of the free layer, $\boldsymbol{m_p}$: spin polarization, both are unit vectors). Therefore, a STFMR signal can only be obtained by either tilting $\boldsymbol{m}$ or $\boldsymbol{m_p}$ (or both). In practice, we apply an in-plane field which then tilts the magnetization orientation of the free layer away from the easy axis. For the initial analysis, the magnetization of the reference layer is assumed to be fixed along the easy axis of the device so that $\boldsymbol{m_p} = (0,0,\pm 1)$. As illustrated in Fig. 1, we can tilt the applied field in our set-up by choosing $H_y \neq 0$.

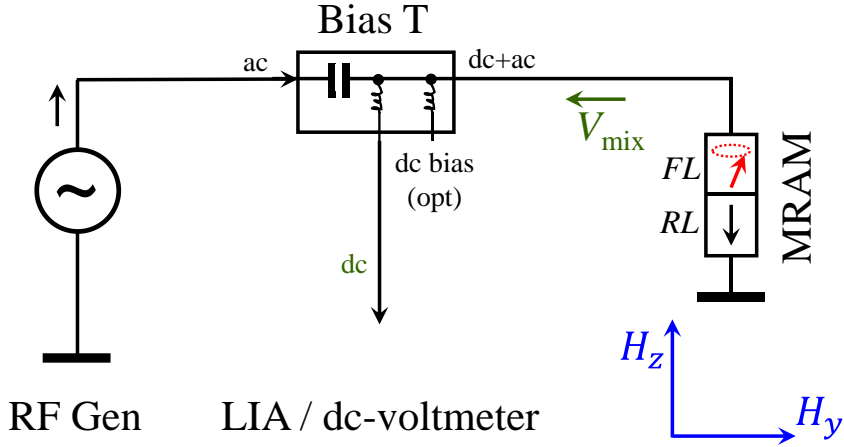

*Fig. 1. Block diagram of the STFMR set-up. The excitation is provided by an RF generator and fed via a Bias T into a MRAM device that is set to the antiparallel state. A perpendicular ($H_z$) and an in-plane ($H_y$) field can be applied independently from one another. LIA=Lock-In Amplifier.*

The perpendicular field $H_z$ and the in-plane field $H_y$ can be controlled independently from one another with maximum strengths of 64 and 110 kA/m, respectively. Frequencies up to 40 GHz are available.

### II.3 Signal-to-Noise Considerations and Calibration

In the following, the basic idea of STFMR is briefly reviewed. The generator drives a current through the device: $I_{RF}(t) = I_{0p} \cos \omega t$. The resistance of the device responds as follows: $R(t) = R_0 + \Delta R \cos \omega t$. Consequently, the voltage across the device is: $V_{mix}(t) = I_{0p} \cos \omega t \, (R_0 + \Delta R \cos \omega t)$. The mixing voltage has components of frequency $\omega$, $2\omega$ and, most importantly, a dc component. The operation of the MRAM is the same as that of a Lock-in amplifier.

Experimentally, $V_{mix}$ is extracted with a Bias T and can, in principle, be measured with a dc voltmeter. In practice, it is advantageous to introduce a modulation where the strength of the excitation is varied and to measure the signal with a Lock-in amplifier rather than a dc voltmeter. Apart from the benefit of ac detection, this effectively becomes a double lock-in detection and offers a signal-to-noise advantage. As pointed out by Goree[18], noise that is caused by RF pickup at the same frequency as the signal itself (here the excitation frequency) is not suppressed by the high frequency Lock-in amplifier (here the MRAM), but can be suppressed by a second, low frequency Lock-in amplifier. The signal-to-noise gain increases with higher modulation frequency but becomes eventually limited by the bandwidth of the low frequency circuit, which is here the dc path of the Bias T (-3dB frequency is 1326 Hz). For our experiments, we use amplitude modulation at 1349 Hz with a modulation depth of 20%. An alternative to amplitude modulation is field modulation[19]. Field modulation can be done by either modulating $H_z$ or $H_y$, and is limited for our set-up to modulation frequencies lower than 40 Hz and therefore not preferred.

It is highly desirable to keep the RF power at the device constant during a measurement. The RF circuitry inevitably involves parasitic capacitances such that the power arriving at the device is reduced with increasing frequency. To calibrate this out, a functional MRAM cell in the antiparallel state (AP) is used. The resistance of the AP state is temperature dependent and can be used for calibration. Starting at low frequency where the capacitive shunting can be neglected, the resistance of the MRAM is measured. When the frequency is increased, less power arrives at the device, its temperature decreases accordingly, and the measured resistance increases. The generator power is then adjusted until the initial resistance at low frequency is restored. In addition, MRAM cells have resistances $\gg 50\Omega$ and cause reflections. Using the same procedure, the effect of the disturbances created by the reflections can be removed by stepping through the frequency range in small steps of 25 MHz.

## III. BASIC STFMR THEORY

Before moving on to the experimental results, it is instructive to derive a simple theoretical result based on a macrospin. Here it is assumed that the magnetization is uniform throughout the cell such that it can be replaced by one single spin. This is justified for small cells where the exchange forces are sufficiently strong, or, for bigger cells, it can be enforced by the application of a high field.

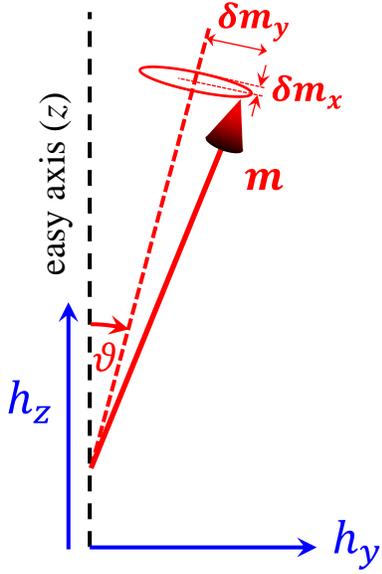

*Fig. 2. Geometry for the macrospin model. The magnetization precesses around its equilibrium orientation as shown with small deviations $\delta m_x$ and $\delta m_y$.*

Normalizing to $\mu_0 M_s H_A' V$, the magnetic energy $e$ is:

$$e = -h_y \sin\vartheta - h_z \cos\vartheta - \frac{1}{2}\cos^2\vartheta$$

(2),

where $\vartheta$ is the angle between the easy axis and magnetization and the lowercase $h$ indicate normalized fields $H/H_A'$. The equilibrium orientation of the magnetization is readily found by solving:

$$\frac{\mathrm{d}e}{\mathrm{d}\vartheta} = -h_y \cos\vartheta + h_z \sin\vartheta + \sin\vartheta \, \cos\vartheta = 0$$

(3).

As indicated in Fig. 2, the magnetization precesses around the equilibrium orientation just calculated. The precession is described by the Landau-Lifshitz equation, which is written in the Gilbert form with the Slonczewski spin transfer torque term added. Here the notation of Butler et al. is adopted[3]:

$$\frac{\mathrm{d}\boldsymbol{m}}{\mathrm{d}t} = \frac{\mu_0 \gamma H_A'}{1+\alpha^2}\left[-(\boldsymbol{m}\times\boldsymbol{h}_{\text{eff}}) - \alpha\boldsymbol{m}\times(\boldsymbol{m}\times\boldsymbol{h}_{\text{eff}}) + \alpha i\left(\alpha+\frac{1}{\eta}\right)(\boldsymbol{m}\times\boldsymbol{m_p})\right.$$
$$\left.+ \alpha i\left(\frac{\alpha}{\eta}-1\right)\boldsymbol{m}\times(\boldsymbol{m}\times\boldsymbol{m_p})\right] = \begin{pmatrix}L_x\\L_y\\L_z\end{pmatrix}$$

$$(4)$$

where $\gamma$ is the gyromagnetic ratio (for our materials $\frac{|\gamma|}{2\pi}$ it has been found to be 29.7 GHz/T rather than the free electron value 28.8 GHz/T), $h_{\text{eff}}$ the effective field acting along the magnetization vector and $i$ the current normalized by $I_{c0}$ as given in equation (1). It is convenient to introduce the natural precession frequency $\Omega_{nat} = \frac{\mu_0 \gamma H_A'}{1+\alpha^2}$. For simplicity, the term $f(I)$ as written in Ref. 3 was replaced by $I$. Considering only small deviations from the equilibrium with subsequent linearization, equation (4) can be written in the frequency domain as (see supplementary material S1):

$$\Omega_{nat}\begin{pmatrix}\alpha h_{\text{stiff2}}+j\omega & h_{\text{stiff1}}\\-h_{\text{stiff2}} & \alpha h_{\text{stiff1}}+j\omega\end{pmatrix}\begin{pmatrix}\Delta\underline{m}_x\\\Delta\underline{m}_y\end{pmatrix} = \alpha\Delta\underline{i}\sin\vartheta\begin{pmatrix}\alpha+\dfrac{1}{\eta}\\\dfrac{\alpha}{\eta}-1\end{pmatrix}$$

$$(5)$$

In equation (5), the two "stiffness fields" $h_{\text{stiff1}} = h_z\cos\vartheta + \cos 2\vartheta + h_y\sin\vartheta$ and $h_{\text{stiff2}} = h_z\cos\vartheta + \cos^2\vartheta + h_y\sin\vartheta$ have been introduced and $\Delta\underline{m}_x$, $\Delta\underline{m}_y$, and $\Delta\underline{i}$ are complex quantities. Note that the excitation is proportional to $\sin\vartheta$, which means that no excitation occurs if the magnetization is on the easy axis. In the zero-damping limit, the determinant of the matrix becomes zero at $\omega^2 = h_{\text{stiff1}}h_{\text{stiff2}}$, which is the classical result from Smit and Beljers[20]. The resonance can then be calculated by solving the two linear equations for $\Delta\underline{m}_x$, and $\Delta\underline{m}_y$.

Finally, the STFMR response is calculated. The resistance is given by:

$$R = \frac{2R_P(1+TMR)}{2+TMR(1+\cos\Theta)}$$

$$(6)$$

where $\Theta$ is the angle between the magnetizations of the reference and the free layers, and $R_P$ is the resistance in the parallel state (P). It is instructive to compare the magnitudes of the resistance change in the P state where $\Theta$ is near 0 and the AP state where $\Theta$ is near $\pi$. Taking the derivative $\mathrm{d}R/\mathrm{d}\Theta$ of equation (6) shows that the same change of $\Theta$ causes a larger resistance change when the MRAM is in the AP state. This is confirmed experimentally and therefore the MRAM is generally set to the AP state as mentioned before. In the following, the considerations are confined to the AP state.

For small excitation, the mixing voltage is (see supplementary material S2):

$$V_{dc,mix} \approx -\frac{R_P TMR(1 + TMR)\sin(\vartheta_{RL} + \vartheta)}{[2 + TMR - TMR\cos(\vartheta_{RL} + \vartheta)]^2} \, \Delta\hat{\imath}\,\Delta\widehat{m}_y\cos\phi = V_0\,\Delta\hat{\imath}\,\Delta\widehat{m}_y\cos\phi$$

(7)

Here $\phi$ is the phase shift of the magnetization relative to the excitation current and $\vartheta_{RL}$ is the equilibrium angle of the magnetization of the reference layer (which is no longer assumed to be infinitely stiff). If the in-plane field $H_y$ is zero, $\vartheta_{RL} = \vartheta = 0$ and the mixing voltage is zero, giving yet another reason as to why an in-plane field $H_y$ is needed. This is evident from Fig 2: if the magnetization precesses around the z-axis, $\cos\Theta$ in equation (6) remains constant and no signal is generated. Consequently, there are two reasons that mandate the presence of an in-plane field for STFMR: a) the ability to excite a resonance at all and b) breaking the symmetry at rectification. Equation (7) indicates that a finite stiffness of the reference layer increases the magnitude of the mixing voltage in the AP state, see supplementary material S2. Combining (6) and (7) gives the final solution and can be written in closed form (note: $V_0 < 0$ in the AP state):

$$V_{dc,mix} \approx \frac{V_0\alpha(h_{\text{stiff2}}(1 + \alpha\eta)\omega^2 - h_{\text{stiff1}}(h_{\text{stiff2}}^2(1 + \alpha^2)^2 + \alpha(\alpha - \eta)\omega^2))\sin\vartheta}{\eta(h_{\text{stiff2}}^2\alpha^2\omega^2 - 2h_{\text{stiff1}}h_{\text{stiff2}}\omega^2 + \omega^4 + h_{\text{stiff1}}^2(h_{\text{stiff2}}^2(1 + \alpha^2)^2 + \alpha^2\omega^2))}\Delta\hat{\imath}$$

(8)

Fig. 3 shows an example of a calculated STFMR response as a function of frequency, normalized to $\Omega_{nat}$. Also shown are the magnitude of $\Delta\widehat{m}_y$ and the phase angle $\phi$. Here the parameters of material A were used for $D = 20$ nm. The STFMR response shows the typical mixture of a Lorentzian and anti-Lorentzian and compares well with experiment. The asymmetrical shape is caused by the phase sensitive term $\cos\phi$ and causes the change in prefix near the resonance. Panel d) shows how the damping can be conveniently extracted from experimental data where $\alpha \approx 2\frac{\omega_B - \omega_A}{\omega_B + \omega_A}$. This is a linewidth construction; however, it is normalized to the resonance frequency and consequently not increase with increasing frequency. It was verified that it successfully retrieves the damping that was input over the range of typical measurement parameters.

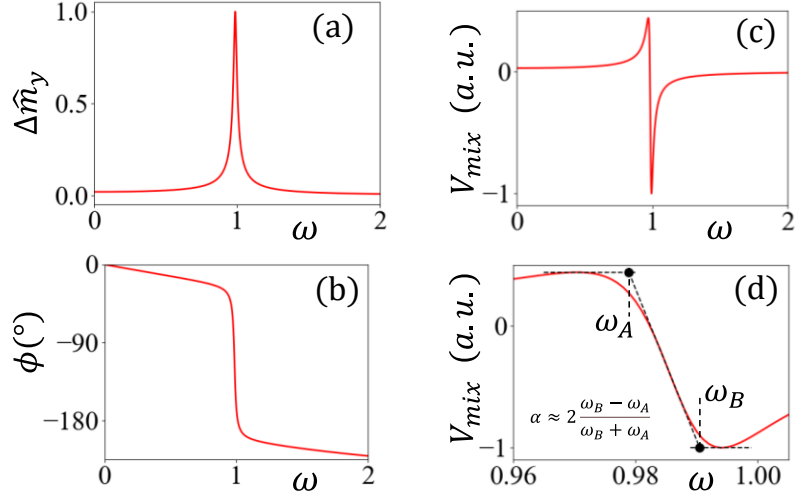

*Fig 3: (a) Amplitude of magnetization component $\Delta\hat{m}_y$ for a macrospin as a function of normalized frequency, where $\omega = 1$ corresponds to the natural precession frequency. Panel (b) shows the corresponding phase response, (c) the mixing voltage as predicted by equation (8), and (d) shows how the damping can be extracted.*

## IV. EXPERIMENTAL RESULTS

Fig. 4 shows an example for a measurement result of a cell with size 26 nm. The size was determined electrically as outlined in[7,8]. For each device, a test scan at $H_z = 0$ and $\pm H_y$ is done, where most of the time the signal is higher for one polarity of $H_y$. Even on the same wafer, this asymmetry differs from device to device, which means that the devices have a natural tilt between either the easy axis and the film normal, or the spin polarization and the film normal, or both. This explains why signals could be seen in[15] without an explicit application of an in-plane field. A typical magnitude of the in-plane field is 10-15% of the total anisotropy field $H'_A$, where generally the polarity is chosen which yields the higher signal.

The best signal quality is achieved by rapidly sweeping $H_z$ back and forth multiple times using a short time constant of 3 ms for the Lock-in amplifier. The sweeps are subsequently averaged, which once again reduces noise that sits on the excitation frequency. Fig.4 (a) gives an isolated trace which was acquired at 21 GHz and Fig.4 (b) shows a contour plot where the resonance is traced over multiple frequencies. It is noted that in standard FMR with zero in-plane field $H_y$, the resonance frequency and the applied field $H_z$ are strictly proportional and are often used interchangeably. For $H_y \neq 0$, equation (5) predicts a small deviation from a straight line that is hardly seen in the field range for $H_z$ investigated here. It is noted, however, that the degree of the curvature seen in Fig. 4 is not predicted by the theory; its origin is the effect of thermal agitation and is discussed in the next section. This additional curvature leads to a very weak dependence (about 1%) of the extracted anisotropy fields on the field $H_z$ and can be largely ignored.

For convenience, the damping is plotted as a function of frequency in the remainder of the paper.

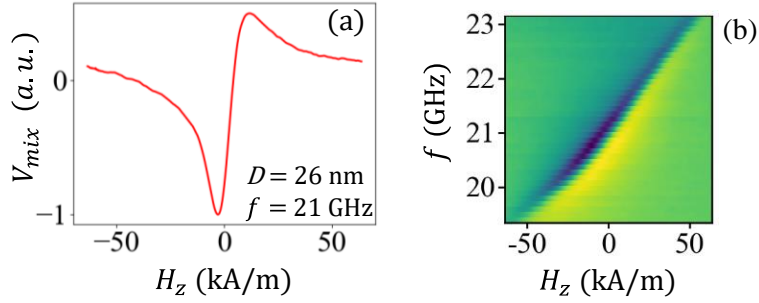

*Fig. 4: Panel (a) shows an isolated STFMR trace as a function of perpendicular field $H_z$ for a 26 nm device at fixed frequency 21GHz. Panel (b) gives a contour plot of the STFMR response across the applied fields and frequencies. Material A.*

Fig. 5 shows a representative result of extracted damping values for multiple sizes for material A. Each point was obtained by evaluating a STFMR trace as shown in Fig. 4(a) according to the scheme discussed in Fig. 3(d) in field rather than frequency space. As expected, the resonance frequencies increase with decreasing size owing to the effect of shape anisotropy. Interestingly, the extracted damping depends on the frequency, or stated more rigorously, on the magnitude of the applied field $H_z$. It is noted that positive $H_z$ points along the magnetization of the free layer (see also Fig. 1) and thus goes along with an increase of resonance frequency. A second trend is that smaller devices show higher damping, where the measured damping is always higher than the film level value $\alpha_0$.

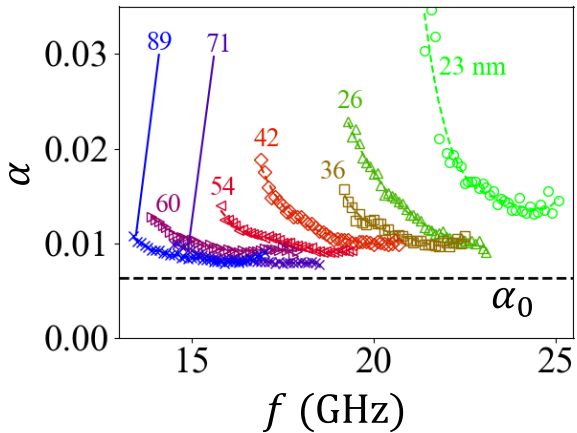

*Fig. 5: Extracted damping for multiple sizes. Each data point was obtained from a field scan $H_z$ at a fixed frequency. The dashed lines are guides to the eye. $\alpha_0$ indicates the film level damping. The numbers are the electrically determined device sizes in nm. Material A.*

Finally, Fig. 6 shows extracted damping data for all three materials across all tested sizes. The behavior is generic: device level damping is found to be higher than the film level damping and increases with decreasing size. For all free layers, the damping is field dependent as seen in Fig. 5. In most cases, the damping has a minimum as a function of the perpendicular field, see for example $D = 23$nm in Fig. 5, but in some cases, for example $D = 26$nm in Fig. 5, the minimum damping appears on the edge of the field range and could potentially be somewhat lower. In Fig. 6, these are marked as filled and open symbols, respectively. These minima have also been reported in[17].

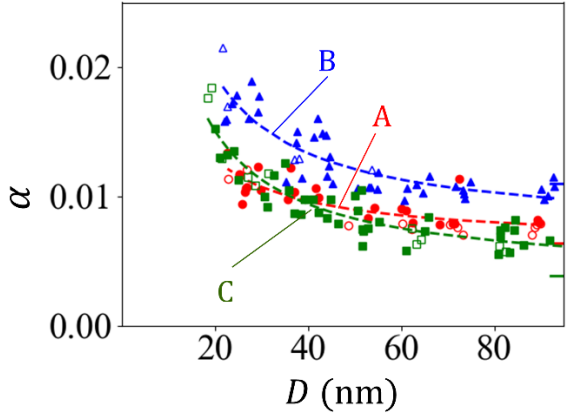

*Fig. 6: Extracted damping for three free layers across multiple sizes. Where present, the damping was extracted from the local minimum (filled symbols), otherwise it was obtained at the edge of the measurement range (open symbols). The markers on the right show the respective film level damping. The dashed lines are guides to the eye.*

## V. EFFECT OF THERMAL ACTIVATION AT FMR RESONANCE

### V.1 Theory

The macrospin model derived in section III reproduces the experimentally observed STFMR response. However, it fails to explain the enhanced damping and in particular, it does not predict the field dependence of the damping. It is plausible that an increase in damping can be caused by edge damage that inevitably occurs when the devices are patterned. A more detailed discussion on this is given further below, but first we want to address the effect of thermal agitation.

Thermal agitation leads to a variety of well-known phenomena: superparamagnetism[21], time dependent coercivity[22], thermal stability[23,24] and magnetic head noise[25], for example. It appears that thermal activation effects can also interfere with ferromagnetic resonance, which, to our knowledge has not been reported before. Considering a 20 nm cell of material A in zero applied field, the rms fluctuation of the polar angle can be estimated to be $\vartheta_{rms} \approx \sqrt{\frac{k_B T}{\mu_0 M_s H_A' V}} = 4.6°$ [26].

For a strongly excited resonance, the corresponding rms value can be estimated from the signal strength to be around 1° and is considerably smaller than that of the thermal excitation.

This effect can be simulated by adding a random field to the LLG equation as suggested by Brown (Langevin dynamics) [27,3]:

$$H_{th} = \frac{1}{\mu_0} \sqrt{\frac{2\alpha k_B T}{M_s V \gamma}} \frac{1}{\sqrt{\Delta t}}$$

(9)

This "thermal field" has a Gaussian distribution with zero mean and standard deviation $H_{th}$. It is considered to be constant for a correlation time $\Delta t$ which is taken to be 1 ps on physical grounds as outlined in[28]. For the analysis, the magnetization vector $\boldsymbol{m}$ is written in polar coordinates $\vartheta, \varphi$ where it is implicitly assumed that its length remains unity. Equation (4) is then:

$$\vartheta'(t) = -\frac{L_z}{\sin\vartheta(t)}$$

(10a)

$$\varphi'(t) = \frac{\cos\varphi(t)L_y - \sin\varphi(t)L_x}{\sin\vartheta(t)}$$

(10b)

where $L_x, L_y$, and $L_z$ were defined in equation (4) and the random field $H_{th}$ must be added to each of the components of the effective field. Fig. 7 (left) shows the resonance at steady state (a) and the transient behavior (b) without thermal agitation; panels (c) and (d) show their counterparts with thermal agitation. It is evident that the effect is strong. Unlike the small angle approximation discussed before, the STFMR signal has to be obtained by plugging the solutions for the magnetization angles $\vartheta(t)$ and $\varphi(t)$ into equation (6) for $R(t)$. The STFMR response is then the time average of the mixing Voltage $V_{mix} = i(t)R(t)$. The calculations are time consuming, because the thermal agitation is strong in comparison to the excitation of the resonance, see also Fig. 7. Each frequency and field combination requires very long averaging times up to several μs to obtain numerically stable results.

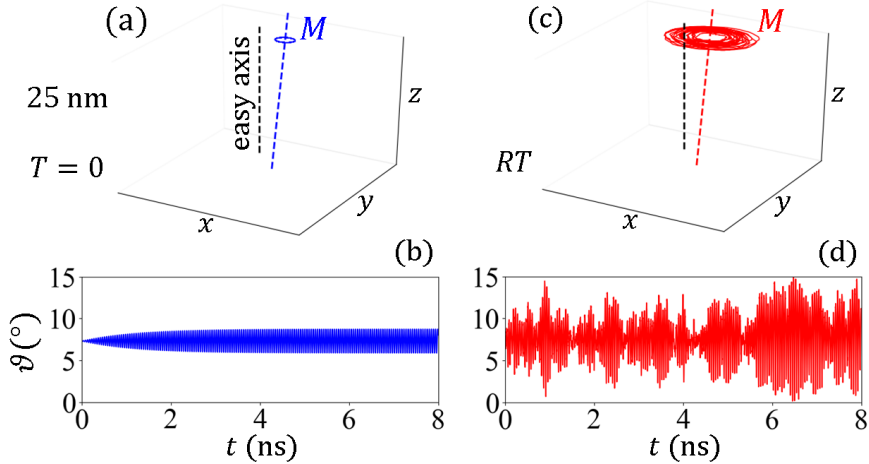

*Fig. 7: Simulation result: polar diagrams (a) and (c) and time traces (b) and (d) for a 25nm cell with free layer A at resonance; with (right) and without (left) thermal agitation, respectively.*

Fig. 8 emulates the experiment and shows the STFMR signal with and without thermal agitation for field scans at fixed frequency for two cell sizes. The frequency is given as a fraction of the respective natural precession frequency. The smaller cell is more affected than the bigger one, as can be seen by the shifts between the curves with and without thermal agitation. One expects that the system becomes stiffer with increasing field such that the same excitation strength creates a smaller resonance signal. This is seen without thermal excitation, but with thermal excitation the increase of the signal strength at negative $H_z$ is partially suppressed and the extracted damping is accordingly increased. The physical reason for this field dependence is that the energy landscape changes with applied field: for negative field, the energy minimum is shallower and, owing to the effect of the thermal energy, the magnetization is on average further away from the minimum.

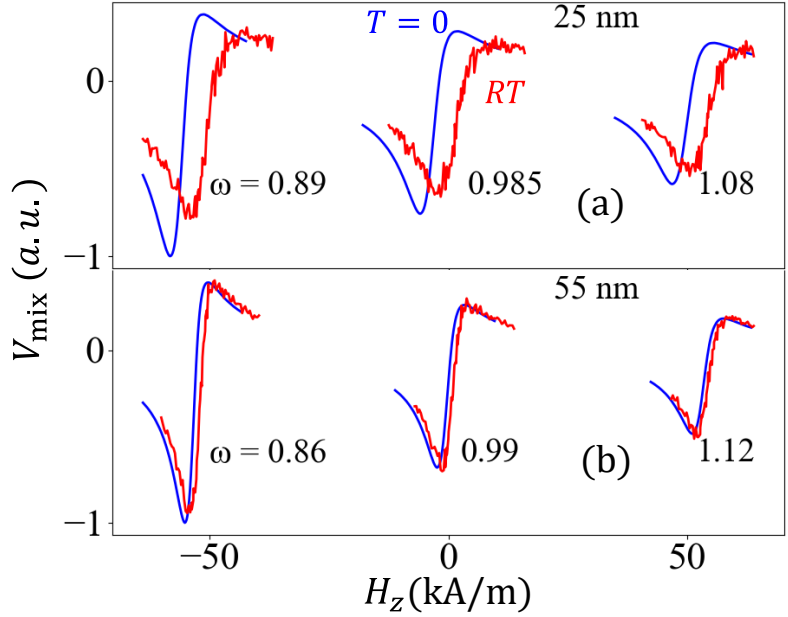

*Fig. 8: Simulated STFMR signals with and without thermal agitation for field scans at fixed frequency for two cell sizes, material A. The frequencies ω are given in fractions of the respective natural precession frequencies.*

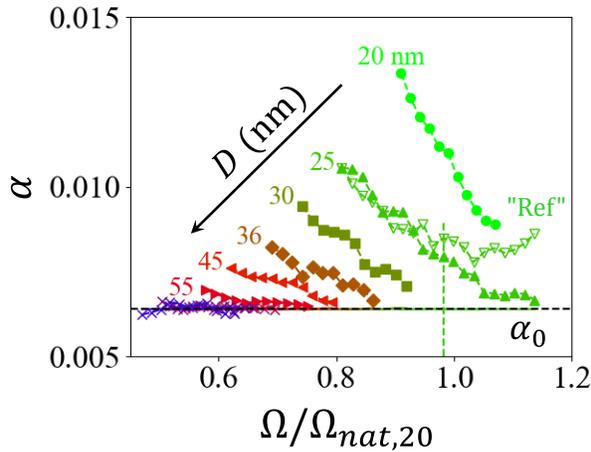

*Fig. 9: Simulated extracted damping for multiple cell sizes including thermal agitation. To replicate the plotting in Fig. 5, the abscissa is normalized to the natural precession frequency of the 20 nm device. To demonstrate the effect of thermal fluctuations of the reference layer, the curves for D=25 nm were extended to higher frequencies, where the filled symbols do not consider thermal fluctuation in the reference layer, as opposed to the open symbols labeled "Ref". All points to the right of the dashed vertical line are not experimentally accessible here. Material A.*

Finally, Fig. 9 gives the simulation result for a series of cell sizes, where the abscissa is normalized to the natural precession frequency of the 20 nm device such that the plot shows the same staggering of the individual curves as the experimental data. For the most part, the experimental data are replicated, and as expected, the intrinsic damping is retrieved for big cells where thermal agitation plays no role. Fitting the damping increase to $e^{-\frac{KV}{k_B T}}$ shows that thermal agitation can be neglected for $KV \gtrsim 262 k_B T$ which corresponds to cell sizes around 55 nm.

A closer look at the experimental data of Fig. 5 shows that the extracted damping tends to go through a minimum and does not show the asymptotic downward trend as simulated. One hypothesis is that also the reference layer is thermally agitated and consequently the excitation current fluctuates too. The measurements are done in the AP state where positive field $H_z$ opposes the magnetization of the reference layer and therefore the fluctuation of the reference layer is *increased* with increasing $H_z$. This was simulated by also subjecting the reference layer to thermal fluctuations, where $KV$ of the reference layer was 1.55 times that of the free layer, its anisotropy field double than that of the free layer and its damping 0.015. To demonstrate the effect, the field and frequency ranges were extended beyond what is accessible experimentally as indicated by the vertical dashed line. As expected, and shown by the filled symbols, the damping approaches the film level value when no thermal agitation of the reference layer is considered. However, if there is a thermal fluctuation of the magnetization of the reference layer, the damping does not reach the film level value and shows a minimum, as seen by the open symbols in Fig. 9, labeled "Ref". It is noted that in experiment, owing to the residual field emanating from the reference layer on the free layer, the curves are shifted a little.

*V.2 Experiment: Temperature Effect*

Varying temperature is an attractive way to check the theory because the amount of thermal agitation can be changed for the *same* device. Fig. 10 shows the result for 25°C, 50°C and 75°C for a 26.4 nm cell. Increasing temperature reduces the total anisotropy field (including shape anisotropy), likewise the natural precession frequency, and consequently all resonance frequencies are expected to be shifted to lower values. For convenience, the figure shows vertical lines where the resonance frequencies are for $H_z = 0$ for the various temperatures. It is noted that these frequencies are close to the natural precession frequencies $\Omega = \Omega_{\text{nat}} \sqrt{1 - \left(\frac{H_y}{H_A'}\right)^2} \approx \Omega_{\text{nat}}$. One can see that the damping increases with increasing temperature as driven by the commensurate reduction of $\frac{KV}{k_B T}$. This conclusion is not altered when the residual stray field from the reference layer (offset field) is considered, because it will merely add to the external field $H_z$ and will shift all three curves by the same small amount. To first order, the resonance frequency follows the perpendicular field, $\Omega \cong \Omega_{\text{nat}}(1 + H_z/H_A')$, and consequently the maximum damping is seen at the lowest frequencies and again, the maximum damping increases with increasing temperature.

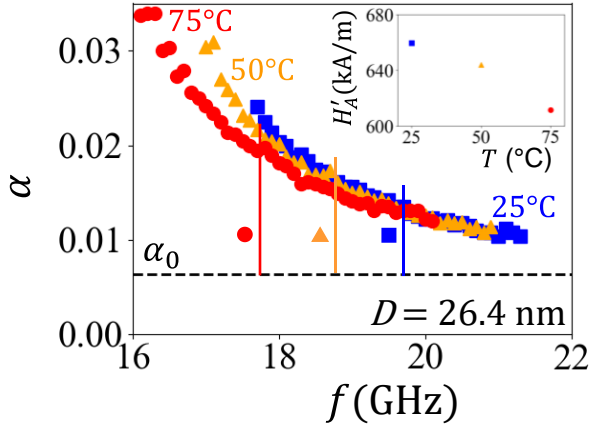

*Fig. 10: Extracted damping for a 26.4 nm device at 3 different temperatures, material A. The inset shows the extracted anisotropy field. The vertical lines indicate where the resonance frequencies are for vanishing $H_z$.*

### V. 3 Damping Increase at Small MRAM Size

The simulations have clearly shown that thermal agitation can increase the *apparent* damping seen in STFMR. On the other hand, the minimum damping values reported in Fig. 6 are higher than the thin film limit, even for the bigger cells with $KV > 262 k_B T$ where the thermal effect can be neglected. It is inevitable that patterning the devices causes edge damage, which is likely to lead to a damping increase. To investigate this further, all data where $KV < 262 k_B T$ were removed and fitted to $\alpha = c_1 + c_2/D$. On physical grounds, any effect on damping caused by edge damage should lead to a scaling of $1/D$, where $c_1$ is expected to agree with the thin film damping $\alpha_0$. The coefficients $c_1$ evaluate to $0.006, 0.0098,$ and $0.0045$ for materials A, B and C, respectively, and, considering the narrow fitting range, are in good agreement with the values given in Table 1.

A question to address is whether increasing the applied field can eliminate the effect of thermal agitation on the extracted damping from STFMR. Considering the free layer alone, as shown in Fig. 9, it is expected that the damping will asymptotically approach the thin film level value for sufficiently strong $H_z$. However, as discussed, the cells are measured in the AP state, and increasing the field destabilizes the magnetization of the reference layer which again causes an increase of the apparent damping. Therefore, it will depend on the details of the cell design whether or not the true damping can be seen for small cells.

### VI. DISCUSSION

In the previous section, it was demonstrated that the apparent damping extracted by STFMR is strongly affected by thermal agitation. The effect becomes negligible for bigger cells with $KV >$ $262k_BT$. However, it appears that the device level damping still is higher than the film level damping. In this section it is discussed what could lead to this behavior.

To investigate whether inhomogeneous magnetization inside the cell can cause increased (apparent) damping, we developed a shell model following a concept given by Andrä and Danan[29]. In this model, different magnetic properties for $H_A$, $M_s$, $A_{ex}$ and $\alpha$ can be assigned to each shell such that edge damage can be investigated. Also, slightly different orientations of the anisotropy within the cell may be considered. A more detailed description of the shell model is beyond the scope of this paper and will be published elsewhere[30], but some of its conclusions are discussed here.

The anisotropy fields extracted from the FMR data are lower than those expected from the film properties for ideal cells but only by about 10% for the smallest cells. This implies that the volume of the damaged material must be small, and the width of a damaged surface layer can at most be of the order of 1 nm. Surface anisotropy effects could also occur at the cylinder surface of the device, similar to the mechanism that gives the cell its perpendicular anisotropy[4]. However, this surface is not well controlled owing to the patterning process and the effect will mostly cancel out. The resulting net effect, if at all present, will lead to a reduced anisotropy near the edges and has the same effect as a reduction of $H_A$ as already considered.

The shell model predicts that these material inhomogeneities do not affect the behavior of the main resonance, because the exchange, even when reduced within reason, is so strong that it effectively couples the magnetization together. When increased damping is assumed for the outermost shell(s), the model successfully retrieves the expected $1/D$ scaling, but it does *not* predict the observed field dependence of the damping. Physically, this means that the shell model justifies an overall increase of the device level damping by introducing a phenomenological increase of the damping at the device edge. The problem of edge damage was also discussed in[31], *albeit* for larger cells $\geq 100$nm with a considerably greater width of the edge damage. There it was found that the damping increased(!) for the larger cells.

In traditional FMR as it has been used to get the film-level damping values here, the damping is extracted from the slope of the line width broadening as a function of field[13] where the field is strictly perpendicular. This "slope $\alpha$" represents the "true" damping. There is a non-zero intercept at zero frequency which is referred to as inhomogeneous line broadening and generally attributed to variations of the anisotropy across the sample. The shell model predicts that the equilibrium magnetization is not strictly uniform throughout the cell. The degree of non-uniformity can be increased by introducing differences in anisotropy between parts of the cell and / or a simultaneous reduction of the exchange. It appears that even for cells as big as 50-60 nm, physically reasonable material inhomogeneities leave the main resonance essentially unchanged. This result suggests that inhomogeneous line broadening does not exist for sizes less than 50-60nm, or stated differently, the anisotropy variations seen in thin film level FMR occur on a larger length scale than that. On the other hand, the shell model predicts the existence of higher order resonance modes. For small cells, such as 20 nm diameter, these occur at high frequencies which are at least twice the main resonance frequency. For these modes, two or more parts of the cell resonate out of phase such that the magnetostatic energy is reduced. These modes are very stiff and consequently the STFMR signal is either too small to be detectable or outside the available frequency range. For bigger cells, the higher order modes move to lower frequencies but even then, they do not interfere with the main resonance. Experimentally, higher order modes, if at all, tend to appear for bigger cells (~60nm and above) and typically have no effect on the extracted damping of the main resonance. This means that the increased damping is not caused by energy transfer between modes.

A well-known mechanism for an increase in damping is *spin-pumping*[32]. Spin pumping occurs when a magnetic and a non-magnetic metal share a common interface so that angular momentum can be transferred from the ferromagnet to the metal. This explains why very thin free layers contacted with a metal electrode show increased damping[1,33]. For our MRAM cells, the free layer has interfaces with MgO in the perpendicular direction, which is known to suppress spin pumping[34]. The circumference of the free layer is in contact with a nonconducting filler. Hence, there is no (intentional) contact to a metal and spin pumping is not expected. However, spin pumping cannot be completely ruled out, because unintentional metallic residues or redepositions could exist on the side walls, which then could lead to a small damping increase.

In conclusion, device level damping measurements across MRAM cell populations of varying size have been found to show increased damping with decreasing cell size. This was observed for three different free layer materials. It has been shown that increased damping is caused by the effect of thermal fluctuations at ferromagnetic resonance. This effect is fundamental and does not reflect a true damping increase. Beyond the thermal effect, we found that there is still an increase of the damping with decreasing size, most likely caused by edge damage during device patterning. The role of these imperfections was theoretically studied by a shell model which justifies an overall increase of the device level damping by introducing a phenomenological increase of the damping at the device edge. Additionally, the shell model predicts that inhomogeneous line broadening does not exist in cells that are smaller than 50 to 60nm.

## VI. SUPPLEMENTARY MATERIAL

Supplementary material S1 and S2 include details how to derive equations (5) and (7), respectively.